\pdfoutput=1
\documentclass[amsmath,amssymb,prl,
showpacs,superscriptaddress,preprintnumbers]{revtex4}

\usepackage{graphicx}
\usepackage{latexsym}
\usepackage{bm}
\usepackage{color}

\begin{document}

\title{Neutrino masses, cosmological bound and four zero Yukawa textures}


\author{Biswajit Adhikary}
\email[Electronic mail: ]{biswajit.adhikary@saha.ac.in}
\affiliation{Saha Institute of Nuclear Physics, 1/AF Bidhannagar, Kolkata 700064, India}

\affiliation{Department of Physics, Gurudas College, Narkeldanga, 
Kolkata 700054, India}

\author{Ambar Ghosal}
\email[Electronic mail: ]{ambar.ghosal@saha.ac.in}
\affiliation{Saha Institute of Nuclear Physics, 1/AF Bidhannagar, Kolkata 700064, India}

\author{Probir Roy}
\email[Electronic mail: ]{probir.roy@saha.ac.in}
\affiliation{Saha Institute of Nuclear Physics, 1/AF Bidhannagar, Kolkata 700064, India}
\vskip 0.50in
\begin{abstract} 
\noindent
Four zero neutrino Yukawa textures in a specified weak basis, combined with 
$\mu\tau$ symmetry and type-I seesaw, yield a highly constrained 
and predictive scheme. 
Two alternately viable $3\times3$ light neutrino Majorana mass matrices 
$m_{\nu A}/m_{\nu B}$ result with inverted/normal mass ordering.  
Neutrino masses, Majorana in character and  
predicted within definite ranges with laboratory and cosmological inputs, 
will have their sum probed cosmologically. 
The rate for $0\nu\beta\beta$ 
decay, though generally below the reach of planned experiments, could 
approach it in some parameter region.
Departure from $\mu\tau$ symmetry due to 
RG evolution from a high scale and consequent CP violation, with a 
Jarlskog invariant whose magnitude could almost reach $6\times 10^{-3}$, 
are explored.
\end{abstract}
\maketitle
Cosmological observations provide a powerful tool 
\cite{one} 
to probe certain properties of elementary particles. 
Such is {\it a fortiori} the case \cite{two,three,tin}
with the sum of the three light neutrino masses $m_1$, $m_2$, $m_3$. 
Its cosmological determination will be of general interest since 
stable light neutrinos comprise hot dark matter in the Universe.   
Moreover, it will pin down the absolute scale of the 
new physics \cite{threeprime} beyond the 
Standard Model already 
discovered from neutrino flavor conversion and oscillation experiments. 
The cosmological upper limit on 
the said mass sum has recently been claimed, by inclusion of data from 
large scale redshift survey, to be  
\cite{four} 
0.28 eV within a $95\%$ c.l. Given the uncertainties 
\cite{five}
due to priors, a value like 0.5 eV is generally accepted 
\cite{six}
now. Further proposed observations 
\cite{seven, sat,eight} 
aim to bring this bound down by more than half an order magnitude. 
KATRIN, an experiment on tritium $\beta$-decay in preparation 
\cite{nine},
aims to probe the absolute neutrino mass scale down to 0.2 eV.  
Meanwhile, 
solar as well as atmospheric, reactor and accelerator 
neutrino and antrineutrino experiments have measured 
\cite{ten}
the mass squared differences 
$|\Delta_{32}^2|\simeq 2.4\times {10}^{-3}$ ${\rm eV}^2$ 
and 
$\Delta_{21}^2\simeq 7.6\times {10}^{-5}$ ${\rm eV}^2$, 
$\Delta_{ij}^2\equiv m_i^2 - m_j^2$, 
 leading to a lower bound 
$\sim$ 0.05 eV on the mass sum. Within an order of magnitude 
(from 0.05 to 0.5 eV) then, 
the sum of the neutrino masses remains undetermined.   
\par
This note is aimed at drawing attention to allowed $\mu\tau$ symmetric 
four zero Yukawa textures 
\cite{eleven}
from which, via type-I seesaw 
\cite{elevenprime},
the above  mass sum as well as individual neutrino masses obtain 
in terms of either of the mass squared differences. 
Such four zero Yukawa textures have a two-fold importance. First, four is the 
maximal number of zeroes phenomenologically allowed 
\cite{seventeen} in the neutrino Yukawa coupling matrix. Second, they 
completely fix \cite{seventeen} the high energy CP violating phases needed 
for leptogenesis in terms of low energy phases observable in the laboratory.
Though the framework used 
here was developed by us earlier 
\cite{eleven},
we now present several new and interesting results. Apart from the expressions 
for the individual neutrino masses, we explicitly calculate the two 
Majorana phases as well as the effective mass $m_{\beta\beta}$ relevant 
to $0\nu\beta\beta$ decay and the Jarlskog invariant $J_{\rm{CP}}$ for 
CP violation. 
We assume only three 
light neutrinos, identical masses for neutrinos and antineutrinos and the 
absence of any new nonstandard interaction. 
The literature today is flush with neutrino mass and mixing models 
\cite{twelve}. However, the idea of $\mu\tau$ symmetric four zero Yukawa 
textures provides a novel angle to view the concerned phenomena and probe the 
mystery of large flavor mixing in the lepton sector. 
This is especially since a texture postulate is more 
basic in a coupling matrix appearing in the 
Lagrangian rather than in the seesaw-derived neutrino mass matrix. 
Another reason for highlighting this model and these calculations is 
that the high scale 
CP violation, needed to describe leptogenesis, is determined \cite{eleven} 
here totally by the low energy CP violation observable in laboratory 
experiments.
\par
We start with a tabulated summary of the esential features of
$\mu\tau$ symmetric four zero Yukawa textures [12]. 
The new results
presented in this communication are explicit expressions for the
neutrino masses $m_1, m_2, m_3$ as well as for the neutrino Majorana
phases and the derivation of numerical bounds on them. Similar results
are also presented on $0\nu\beta\beta$ decay and CP violation.
In a nutshell, the three neutrino masses in our scheme may be written as 
$$
m_{1,2} = {\left|\Delta_{21}^2\left(\frac{2-X_3\mp X}{2X}\right)\right|}^{1/2},
m_3  = \left|\Delta_{21}^2/X\right|^{1/2}.
\eqno(1)
$$
(The masses can be rewritten in terms of $|\Delta_{32}^2|$, 
$X$ and $X_3$.)
Here $X$ and $X_3$ are dimensionless parameters which are numerically 
restricted by neutrino oscillation data. With that and 
the earlier mentioned inputs, the predicted 
interval for 
$m_1 + m_2 + m_3$ (using $3\sigma$ oscillation data) is [0.156, 0.5] eV/ [0.074, 0.132] eV
for inverted/normal mass ordering with     
$(\Delta_{32}^2<0)$/$(\Delta_{32}^2>0)$. 
\par
There appears to be something magical about four zero Yukawa textures. Their 
phenomenological success vis-a-vis up and down type quark Yukawas is 
well-documented 
\cite{thirteen,tero, arotero,fourteen}.
In the lepton sector it is convenient to work 
\cite{fifteen} 
in a flavor basis with mass diagonal charged leptons $l_{\alpha}$ 
$(\alpha = e,\mu,\tau)$. Within type-I seesaw, 
that weak basis gets completely specified by the additional choice 
\cite{seventeen}
of three mass diagonal heavy right chiral neutrinos $N_i$ ($i = 1,2,3$) with
real masses $M_i$. Now four has been demonstrated 
\cite{seventeen}
to be the maximum number of texture zeroes allowed in the neutrino 
Yukawa coupling matrix $Y_\nu$ in flavor space -- given 
the requirement of three lightly massive neutrinos and no unmixed neutrino flavor. High scale 
CP violation, as required for leptogenesis, then gets completely determined 
\cite{seventeen}
in terms of phases that are in principle measurable in laboratory 
experiments.
\par
Such a scheme of four zero $Y_\nu$ textures is 
both strongly constrained 
\cite{eighteen,merle} by neutrino oscillation data and is at the same time 
quite 
predictive regarding radiative charged lepton decays. The additional 
introduction 
of $\mu\tau$ symmetry 
\cite{eleven}, 
motivated by the observed near maximality of $\nu_\mu$-$\nu_\tau$ 
mixing, drastically reduces the number of the 
allowed textures from seventy two 
to four. The four allowed textures fall pairwise 
into 
two categories $A$ and $B$ with a single light neutrino Majorana mass 
matrix ${\cal M}_\nu$ for each category. Apart from one nontrivial 
phase $\bar\alpha$/$\bar\beta$, 
and an overall mass factor $m_A$/$m_B$,  
${\cal M}_{\nu A}$/${\cal M}_{\nu B}$ 
is characterized by two 
real parameters $k_{1,2}/l_{1,2}$. 
The same phase contributes to $0\nu\beta\beta$ decay, leptogenesis 
\cite{eleven,eighteen,nineteen}
as well as to CP violation to be studied in long baseline experiments 
 -- the latter with the introduction of small departures 
\cite{eleven} 
from $\mu\tau$ symmetry.    
\par
We choose to work within the Minimal Supersymmetric Standard Model (MSSM 
\cite{twenty})
 wherein the neutrino Dirac mass matrix is $m_D = Y_\nu v_u/\sqrt{2}$ with 
$v_u\simeq (246\, {\rm GeV})\sin\beta$. 
We have $(Y_\nu)_{12}$ = $(Y_\nu)_{13}$,
$(Y_\nu)_{21}$ = $(Y_\nu)_{31}$,
$(Y_\nu)_{23}$ = $(Y_\nu)_{32}$,
$(Y_\nu)_{22}$ = $(Y_\nu)_{33}$ and 
$M_2=M_3$ from $\mu\tau$ symmetry. The type-I seesaw relation 
${\cal M}_\nu=-m_D\, {\rm diag}(M_1^{-1}, M_2^{-1}, M_3^{-1})\, m_D^T$ 
then yields a $\mu\tau$ symmetric ${\cal M}_\nu={\cal M}_\nu^T$, 
leading immediately to 
$\theta_{23} = \pi/4$, $\theta_{13}=0$. It further follows that 
$H\equiv  {\cal M}_\nu {\cal M}_\nu^\dagger$ 
is also $\mu\tau$ symmetric. Details of these results are available 
elsewhere 
\cite{eleven}.
Suffice it to summarize the relevant parts in TABLE I. 
Here $a_{1,2}$, $b_{1,2}$ are complex parameters. They appear in 
$m_{DA}/m_{DB}$ reappearing in 
${\cal M}_{\nu A}/{\cal M}_{\nu B}$ 
in terms of the real parameters $k_{1,2}/l_{1,2}$ 
and the phases $\bar\alpha/\bar\beta$ \cite{new}.
Furthermore, 
$X_{aA}/X_{aB}$, with $a = 1$ through 5 are explicitly constructed functions 
of $(k_1,k_2,\bar\alpha)$/ $(l_1,l_2,\bar\beta)$ while 
$X$ equals ${(X_1^2+X_2^2)}^{1/2}$. 
\begin{table}
\begin{tabular}{|l|}
\hline
Category $A$\\
\hline
$
m_{DA}^{(1)} = \begin{pmatrix}a_1&a_2&a_2\\
                         0&0&b_1\\
                         0&b_1&0
               \end{pmatrix},\, 
m_{DA}^{(2)} = \begin{pmatrix}a_1&a_2&a_2\\
                         0&b_1&0\\
                         0&0&b_1
               \end{pmatrix}
$\\
\hline
$
m_A = -b_1^2/M_2, 
k_1={\left|{\frac{a_1}{b_1}}\right|}
\sqrt{\frac{M_2}{M_1}},k_2 = \left|{\frac{a_2}{b_1}}\right|, 
\bar{\alpha} = {\rm arg}
\frac{a_1}{a_2}
$\\
\hline
$
{\cal M}_{\nu A} = m_A\begin{pmatrix}{k_1}^2e^{2i\bar\alpha}+2{k_2}^2&k_2&k_2\\
                        k_2 &1& 0\\
                        k_2&0&1\end{pmatrix}
$ \\
\hline
$
X_{1A} = 2\sqrt{2}k_2{[{(1+2k_2^2)}^2 + k_1^4 + 2k_1^2(1+2k_2^2)\cos2\bar
\alpha]}^{1/2} 
$\\
$
X_{2A} = 1-k_1^4-4k_2^4-4k_1^2k_2^2\cos2\bar\alpha
$\\
$
X_{3A} = 1-4k_2^4-k_1^4-4k_1^2k_2^2\cos2\bar\alpha - 4k_2^2
$\\
$
X_{4A}= k_1^4+4k_2^4+4k_1^2k_2^2\cos2\bar\alpha
$\\
$
X_{5A} = -8k_1^2k_2^4\sin2\bar\alpha 
$\\
\hline
Category $B$\\
\hline
$
m_{DB}^{(1)} = \begin{pmatrix}a_1&0&0\\
                        b_1&0&b_2\\
                        b_1&b_2&0\end{pmatrix},\, 
m_{DB}^{(2)} = \begin{pmatrix}a_1&0&0\\
                        b_1&b_2&0\\
                        b_1&0&b_2
                  \end{pmatrix}
$\\
\hline
$
m_B = -b_2^2/M_2, l_1=\left|{\frac{a_1}{b_2}}\right|
\sqrt{\frac{M_2}{M_1}},l_2 = \left|{\frac{b_1}{b_2}}\right|
\sqrt{\frac{M_2}{M_1}},
\bar{\beta} = {\rm arg}\frac{b_1}{b_2}
$
\\
\hline
$
{\cal M}_{\nu B} = m_B \begin{pmatrix}
             l_1^2&l_1l_2e^{i\bar\beta}&l_1l_2e^{i\bar\beta}\\
                                    l_1l_2e^{i\bar\beta}&l_2^2e^{2i\bar\beta}+
1&l_2^2e^{2i\bar\beta}\\
                                    l_1l_2e^{i\bar\beta}&l_2^2e^{2i\bar\beta}
&l_2^2e^{2i\bar\beta}+1
\end{pmatrix}
$\\
\hline
$
X_{1B} = 2\sqrt{2}l_1l_2{[{(l_1^2+2l_2^2)}^2 + 
1+ 2(l_1^2+2l_2^2)\cos2\bar\beta]}^{1/2}
$\\
$
X_{2B} = 1+4l_2^2\cos2\bar\beta+4l_2^4-l_1^4
$\\
$
X_{3B} = 1-{(l_1^2+2l_2^2)}^2 - 4l_2^2\cos2\bar\beta
$\\
$
X_{4B} = l_1^4
$\\
$
X_{5B} = -8l_1^2l_2^4\sin2\bar\beta 
$\\
\hline
\end{tabular}
\caption{Relevant matrices as well as parameters and their functions 
in either category}
\end{table}
Their expressions in terms of the parameters $k_1$, $k_2$,  
$\bar\alpha$ (Category $A$) or $l_1$, $l_2$, $\bar\beta$
(Category $B$) are useful because the $\{X_a\}$ are directly related to 
experimentally measurable quantities such as the mass squared differences, 
mixing angles, effective majorana mass as measured in $0\nu\beta\beta$ decay 
and the CP violating Jarlskog invariant. Specifically, $X$ appears in 
$\Delta_{21}^2$ and $X_3$--$X$ in $\Delta_{32}^2$.
\noindent
Computed eigenvalues of $H$, namely 
the squares of the light neutrino masses, lead to (1). 
Moreover, we have
$$
\Delta_{21}^2 = {|m|}^2 X,
\Delta_{32}^2 = \frac{{|m|}^2}{2}(X_3 - X),
\theta_{12} = \frac{1}{2}\tan^{-1}\frac{X_1}{X_2}.
\eqno(2)
$$
In (2) $m$ is the overall (complex)  mass factor denoted in Table I by 
$m_A/m_B$ for Category $A/B$ and $|m|$ turns out to equal 
$m_3$ in either category. 
\par
We use the following 
\cite{twentyone} $3\sigma$ intervals 
for $\tan2\theta_{12}$, $\Delta_{21}^2$ and 
$\Delta_{32}^2:1.83\leq\tan2\theta_{12}\le 4.90$, 
$7.07\times 10^{-5}\, {\rm eV^2}\le\Delta_{21}^2\le 8.34\times10^{-5}\,
{\rm eV^2}$ 
and 
$2.13\times 10^{-3} \,{\rm eV^2}\le \Delta_{32}^2\le 2.88\times10^{-3}\,
{\rm eV^2}$
\,(normal mass ordering),
$-2.79\times 10^{-3}\, {\rm eV^2}\le \Delta_{32}^2\le -2.02\times10^{-3}\,
{\rm eV^2}$\, (inverted mass ordering).
These permit \cite{nineteen} only inverted (normal) ordering 
for Category $A$ $(B)$ and allow only a constrained domain  in the 
$k_1$-$k_2$/$l_1$-$l_2$ plane. They further restrict \cite{nineteen} 
the magnitude of the 
phases to $89.0^o<|\bar\alpha|<90^o$, $87.0^o<|\bar\beta|<90^o$, allowing  
a limited region in the $X_3$--$X$ plane.
The consequent correlated values of $m_i$ are shown in the top panel plots of 
FIG.1. The limits on the sum of the neutrino masses, 
quoted at the begining, follow as a result. 
For the individual masses $m_i/(\rm eV)$, $i=1,2,3$, we have the respective 
intervals $[0.0452,~0.1682]$, $[0.0457,~0.1684]$,
$[0.0077,~0.1623]$ for Category $A$ and  
$[0.0110,~0.0335]$, $[0.0144,~0.0345]$,  
$[0.0485,~0.0638]$ for Category $B$.
\noindent 
Given the allowed ranges, 
it is not possible right now to distinguish between the 
hierarchical and quasidegenerate possibilities. But a future
reduction of the ranges could pin this down.
\par
One can now calculate the neutrino 
Majorana phases \cite{Adhikary:2006rf} from the explicit diagonalization of 
${\cal M}_\nu$. Let us parametrize the Majorana phase factor 
\cite{twentysix} 
of the neutrino mixing matrix as 
${\rm diag}.(e^{\frac{i\alpha_{{\rm M}_1}}{2}}, 
e^{\frac{i\alpha_{{\rm M}_2}}{2}},1)$. 
We then obtain
$$
\cos(\alpha_{M_1}-arg.Z) = 
\frac{|Z|^2m_3^2+m_1^2\sin^4\theta_{12}-
m_2^2\cos^4\theta_{12}}{
2m_1 m_3\sin^2\theta_{12}|Z|}
\eqno(3a)
$$
$$
\cos(\alpha_{M_2}-arg.Z) = 
\frac{
|Z|^2m_3^2+m_2^2\cos^4\theta_{12}-
m_1^2\sin^4\theta_{12}}{
2m_2 m_3\cos^2\theta_{12}|Z|}
\eqno(3b)
$$
\noindent
with 
$Z = {[{({\cal M}_\nu)}_{22}+{({\cal M}_\nu)}_{23}]}
{[{({\cal M}_\nu)}_{22}-{({\cal M}_\nu)}_{23}]}^{-1}$. 
Restricting $\alpha_{M_1}$, $\alpha_{M_2}$ to the 
$-\pi$ to $\pi$ interval,  
utilizing the expressions for $Z$, $m_1$, $m_2$, $m_3$ and $\theta_{12}$ in 
terms of the basic parameters 
($k_1$,$k_2$, $\bar\alpha$)/($l_1$,$l_2$, $\bar\beta$) and utilizing the 
phenomenologically acceptable ranges of these parameters as given in 
Ref. [14], we numerically find from eqs.(3) 
$-98.0^o\leq \alpha_{M_1}\leq 20.0^o$, 
$9.2^o\leq \alpha_{M_2}\leq 36.4^o$ 
for Category $A$ and 
$-88.6^o\leq \alpha_{M_1}\leq 7.97^o$, 
$90.7^o\leq \alpha_{M_2}\leq 128.8^o$ 
for Category $B$. Allowed values of $\alpha_{M_2}$ vs $\alpha_{M_1}$ are
shown in the middle panels of FIG.1.
\begin{figure}[htb]
\begin{flushleft}
\includegraphics[width=3.5cm,height=3.5cm]{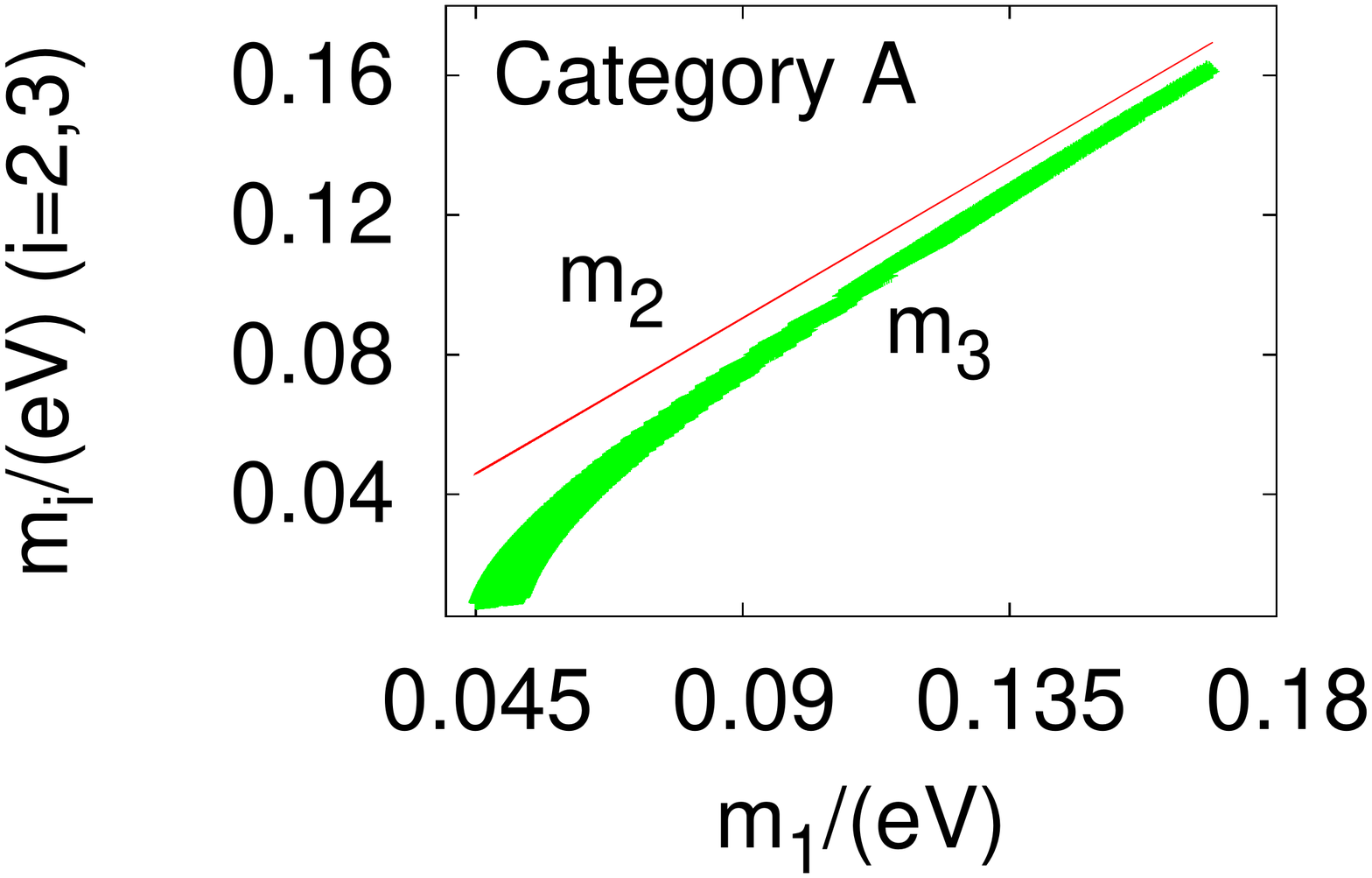}
\includegraphics[width=3.5cm,height=3.5cm]{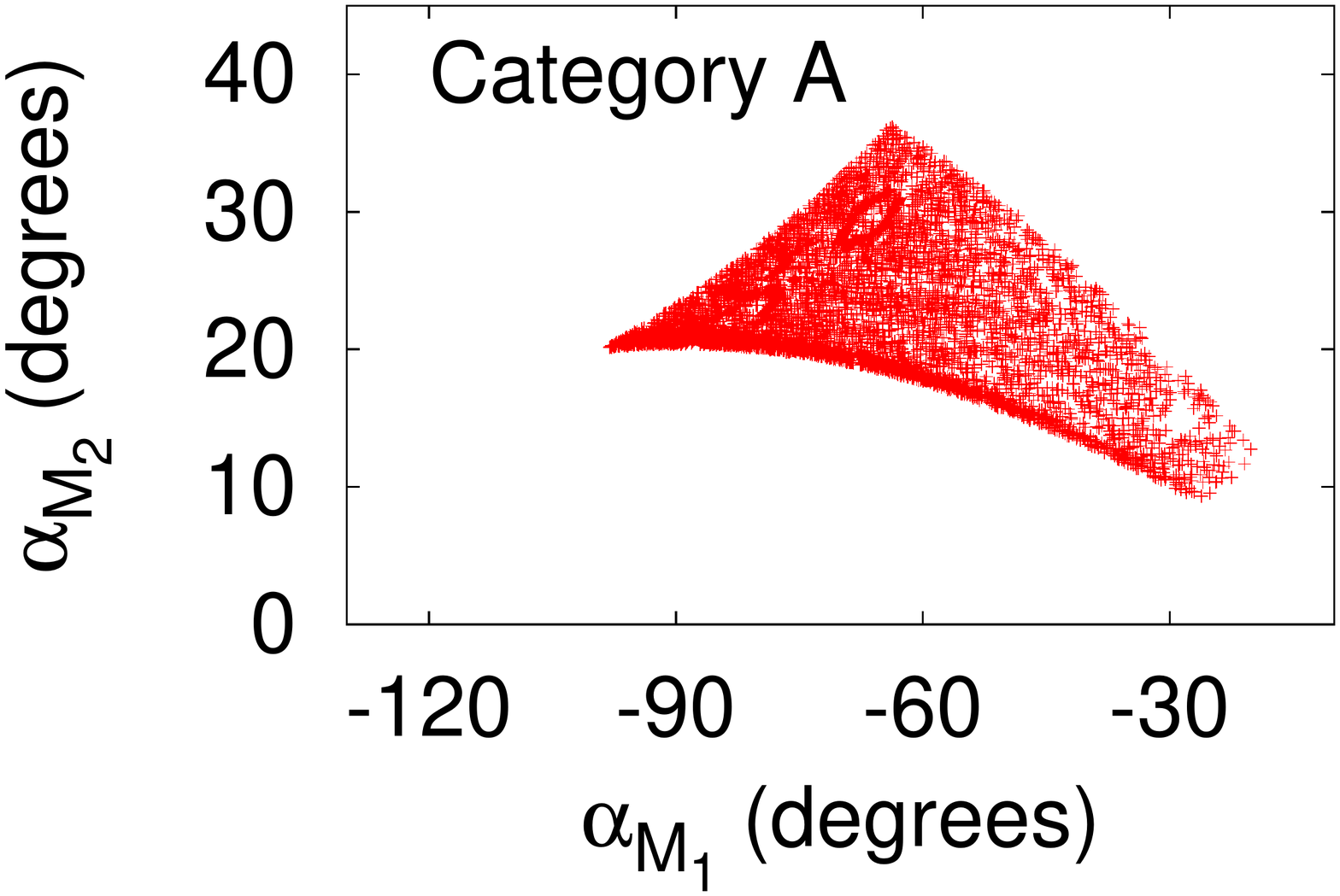}
\includegraphics[width=3.5cm,height=3.5cm]{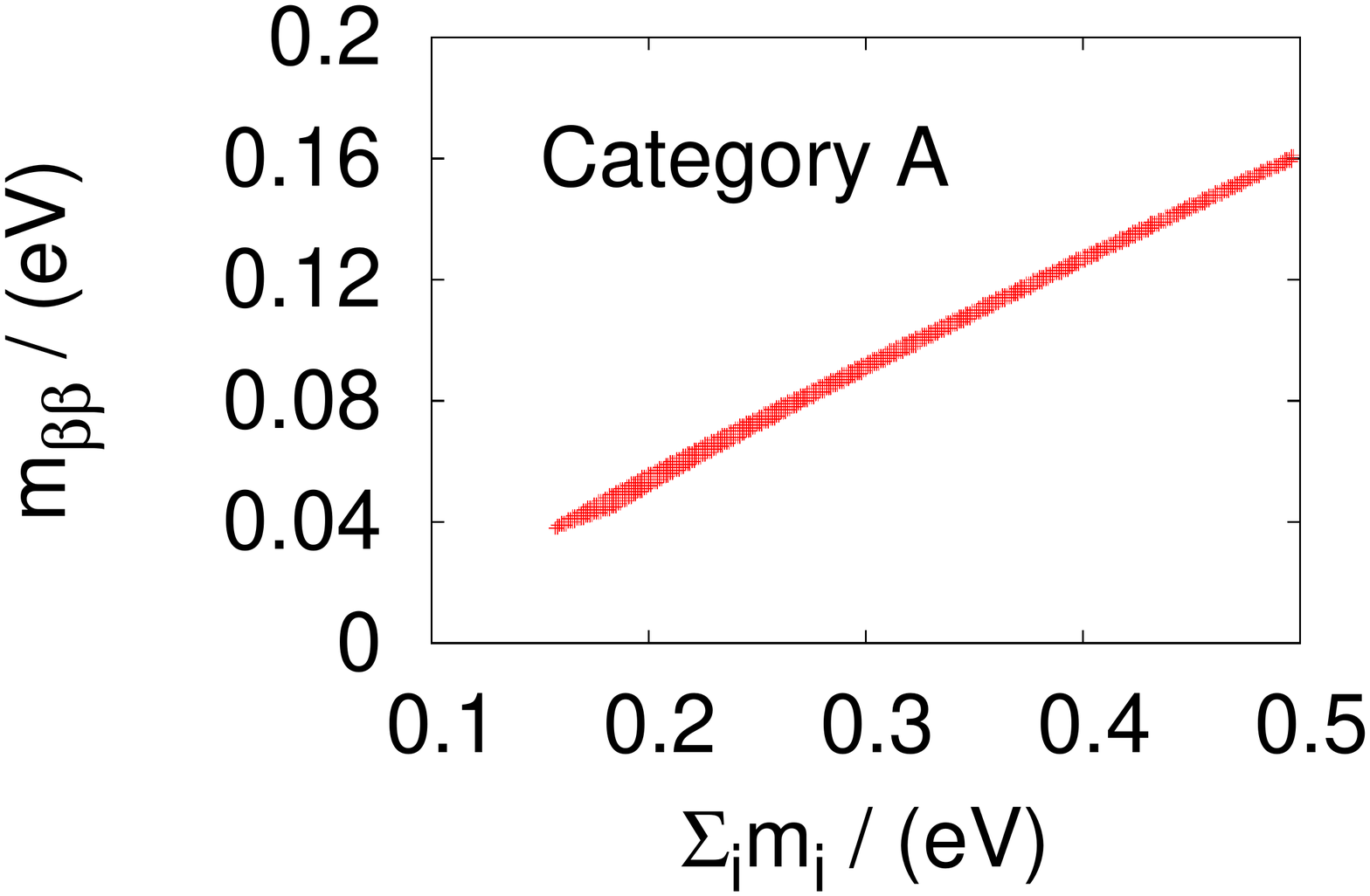}
\vskip 0.1cm
\includegraphics[width=3.5cm,height=3.5cm]{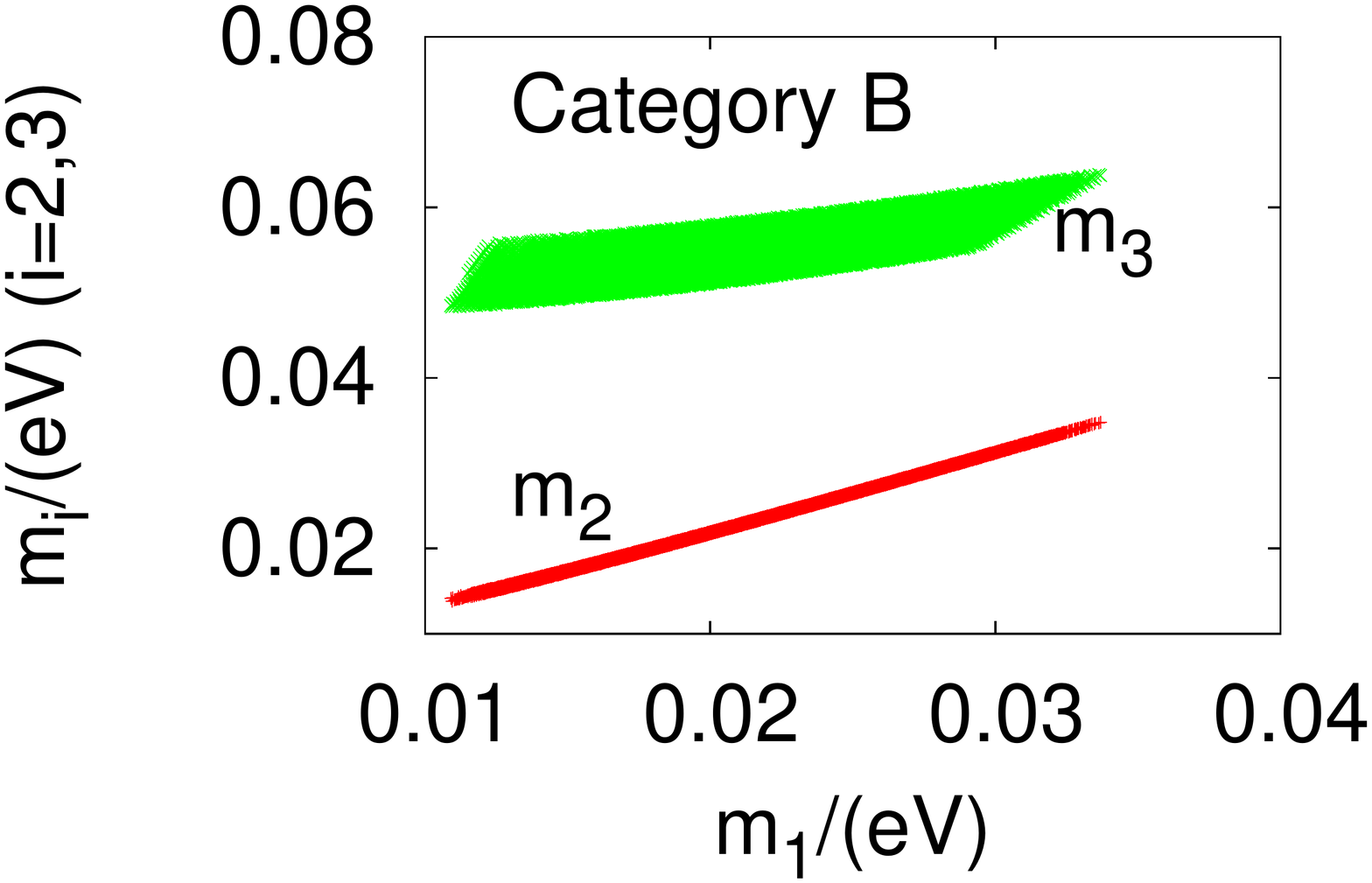}
\includegraphics[width=3.5cm,height=3.5cm]{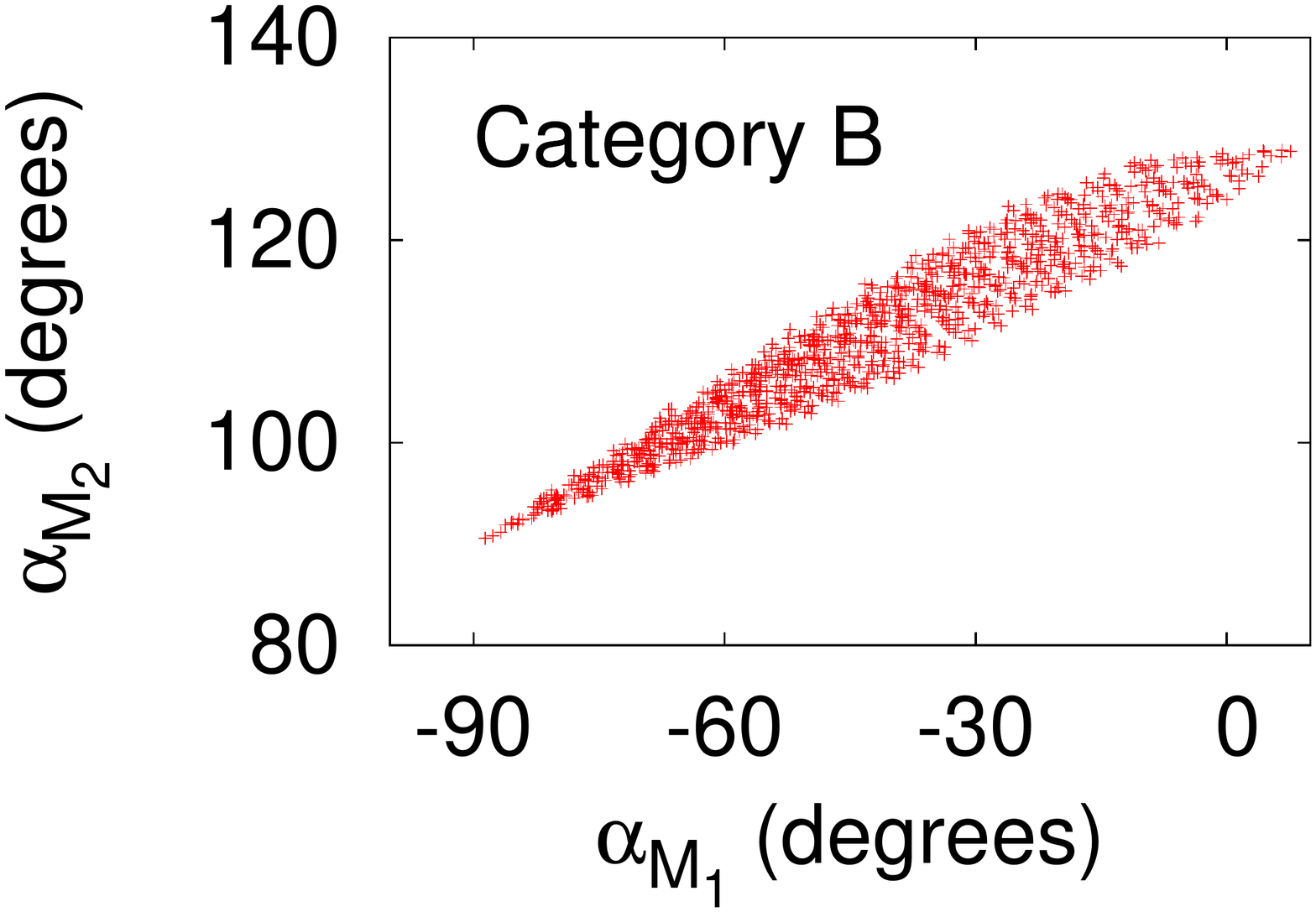}
\includegraphics[width=3.5cm,height=3.5cm]{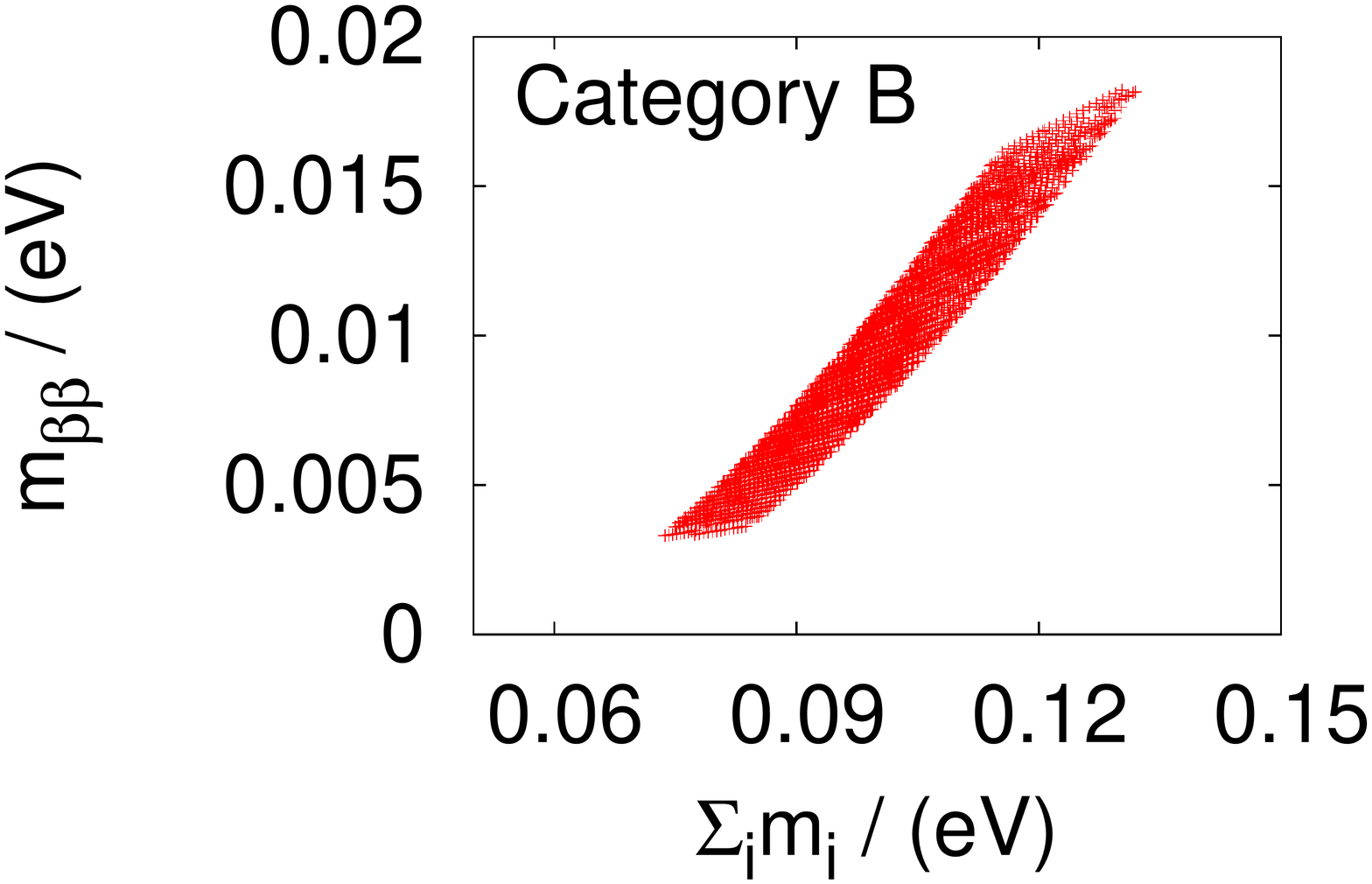}
\caption{\small{{(Color online) Allowed $m_{2,3}$ vs $m_1$ (left), 
$\alpha_{M_2}$ vs $\alpha_{M_1}$ (middle) and  
$m_{\beta\beta}$ vs $\Sigma m_i$ (right) for Category $A$ 
(top) and Category $B$ (bottom).}}}
\end{flushleft}
\end{figure}
\par
The effective mass $m_{\beta\beta} = |{({\cal M}_{\nu})}_{11}|$, appearing in the 
transition amplitude for the yet unobserved neutrinoless nuclear double 
$\beta$-decay, is given in our scheme by
$$
m_{\beta\beta} = {|\Delta_{21}^2 X_4 X^{-1}|}^{1/2},
\eqno(4)
$$
\noindent
with $X_4$ as appearing in TABLE I. Including uncertainties in the relevant 
nuclear matrix element, the currently accepted experimental upper bound 
\cite{twentythree}
is $m_{\beta\beta}<0.35$ eV. In comparison, the predicted 3$\sigma$ 
ranges for $m_{\beta\beta}$ in our scheme 
are (cf. FIG.1) 
$0.038\leq m_{\beta\beta}/{\rm eV}\leq 0.161$ for Category $A$
and 
$0.003\leq m_{\beta\beta}/{\rm eV}\leq 0.0186$
for Category $B$.
Though generally below the planned reach 
\cite{twentyfour, 24}
of the forthcoming searches for $0\nu\beta\beta$ decay, our $m_{\beta\beta}$ 
could approach 
it for Category $A$. It could be seriously probed in 
the subsequent round of experiments. Allowed values of $m_{\beta\beta}$ vs 
$m_1+m_2+m_3$ are shown in the lowermost panels of FIG.1.
\par
There could be deviation from $\mu\tau$ symmetry and 
the values $\theta_{23} = \pi/4$, $\theta_{13} = 0$ 
might not stand up to more precise forthcoming measurements. Indeed, there is 
already a hint 
\cite{ten,twentyone}
of a nonzero $\theta_{13}$ from global analyses of neutrino oscillation data.
We had earlier proposed 
\cite{eleven}
a theory of such a departure based on Renormalization Group (RG) evolution 
from a high energy scale $\Lambda\sim {(M_i)}_{max}\sim {10}^{12}$ GeV, 
where $\mu\tau$ symmetry is deemed exact, to a laboratory scale 
$\lambda\sim {10}^3$ GeV. The said departure is then caused at the 1-loop 
level by charged lepton mass terms. Neglecting $m_{\mu,e}^2$ in comparison 
with $m_\tau^2$, the deviation from $\mu\tau$ symmetry gets characterized by 
a parameter 
\cite{eeleven} retained to linear order and given by 
$$
\Delta_\tau \simeq 
\frac{m_\tau^2}{8\pi^2v^2}(1+\tan^2\beta){\rm ln}\frac{\Lambda}{\lambda}
\leq 0.05,
\eqno(5)
$$
\noindent  
the upper bound being reached for \cite{twenty} $\tan\beta = 60$. 
\par
It now follows that
\cite{eleven}
$$
{\cal M}_{\nu}^\lambda \simeq {\rm diag} (1,1,1-\Delta_\tau)
{\cal M}_\nu^\Lambda {\rm diag} (1,1,1-\Delta_\tau), 
\eqno(6)
$$
where ${\cal M}_\nu^\Lambda$ equals ${\cal M}_\nu$ of TABLE I 
and henceforth we drop all superscripts. 
\noindent
Detailed expressions to $O(\Delta_\tau)$ for the 
two neutrino mass squared differences 
$\Delta_{21}^2$ and $\Delta_{32}^2$ as well as for 
the mixing angles $\theta_{ij}$ were given in the Appendix of 
Ref.~\cite{eleven}. 
Numerical shifts in neutrino masses and the mass sum are minor, 
but now  $0.0098\leq m_{\beta\beta}/{\rm eV}\leq 0.163$ ($0.006\leq 
m_{\beta\beta}/{\rm eV}\leq 0.0224$) for Category $A$ ($B$). A 
nontrivial statement also emerges on the two mixing angles kept free.
In Category 
$A$ ($B$), we have  
$36.3^o\leq\theta_{23}<45^o$ $(45^o< \theta_{23}\leq 
46.8^o)$ and 
$0<\theta_{13}\leq 2.3^o\, (0.84^o)$; see the top panels of FIG.2. 
\par
Turning to the question of CP 
violation, which can be observed 
\cite{twentyfive}
via a long baseline measurement of the difference in oscillation probabilities 
$P(\nu_\mu\rightarrow \nu_e) - P(\bar{\nu_\mu}\rightarrow \bar{\nu_e})$, 
our statement on the magnitude of the PMNS phase $\delta$ is 
$16.0^o\leq |\delta|\leq 89.50^o$\, 
($1.50^o\leq| \delta|\leq 88.05^o$) 
for Category $A$ ($B$). For the basis independent 
Jarlskog invariant, we have 
$$
J_{\rm CP} = {\rm Im}\frac{H_{12} H_{23} H_{31}}{
\Delta_{21}^2 \Delta_{32}^2 \Delta_{31}^2} \simeq 
\frac{X_5\Delta_\tau}{X(X_3^2-X^2)} + O(\Delta_\tau^2).
\eqno(7)
$$
\noindent
$X$, $X_3$ in (7) are as before and $X_5$ for either category has been given 
in TABLE I.
We find that 
$0 < |X_5| \leq 0.203$ 
( $0 < |X_5| \leq 0.0011$), so that 
$0 < |J_{\rm CP}|\leq 0.0086$ 
($0 < |J_{\rm CP}| \leq 0.0026$) 
for Category $A$ ($B$). See the lower panels of 
FIG. 2 for allowed $|\delta|$ vs $|J_{CP}|$. 
The single phase factor $\sin\bar\alpha$/$\sin\bar\beta$, appearing 
$X_{5A}$/$X_{5B}$, is what also shows up 
\cite{nineteen}
in the baryon asymmetry of the Universe originating via high scale 
leptogenesis. This is a concrete realization of the earlier statement 
\cite{seventeen}
that here CP violation -- as 
required for high scale leptogenesis -- is determined by that showing up 
in laboratory energy neutrino oscillations.
The sign of $\delta$ and hence that of $J_{\rm CP}$ 
 get related via the baryon asymmetry to the 
mass hierarchy in $N_i$ [23].    
\par
Let us also address the question of a possible dynamical origin of our 
scheme. While $\mu\tau$ symmetry is 
likely to be a discrete subgroup of a broken flavor symmetry group, 
the real issue is that of the postulated four zero Yukawa textures. In the 
Frogatt-Nielsen approach 
\cite{30}, one 
can regard texture zeroes as extremely suppressed entries which can be taken 
to vanish. 
On the other hand, a general set of Abelian discrete symmetries 
\begin{figure}[htb]
\begin{flushleft}
\includegraphics[width=3.5cm,height=3.5cm]{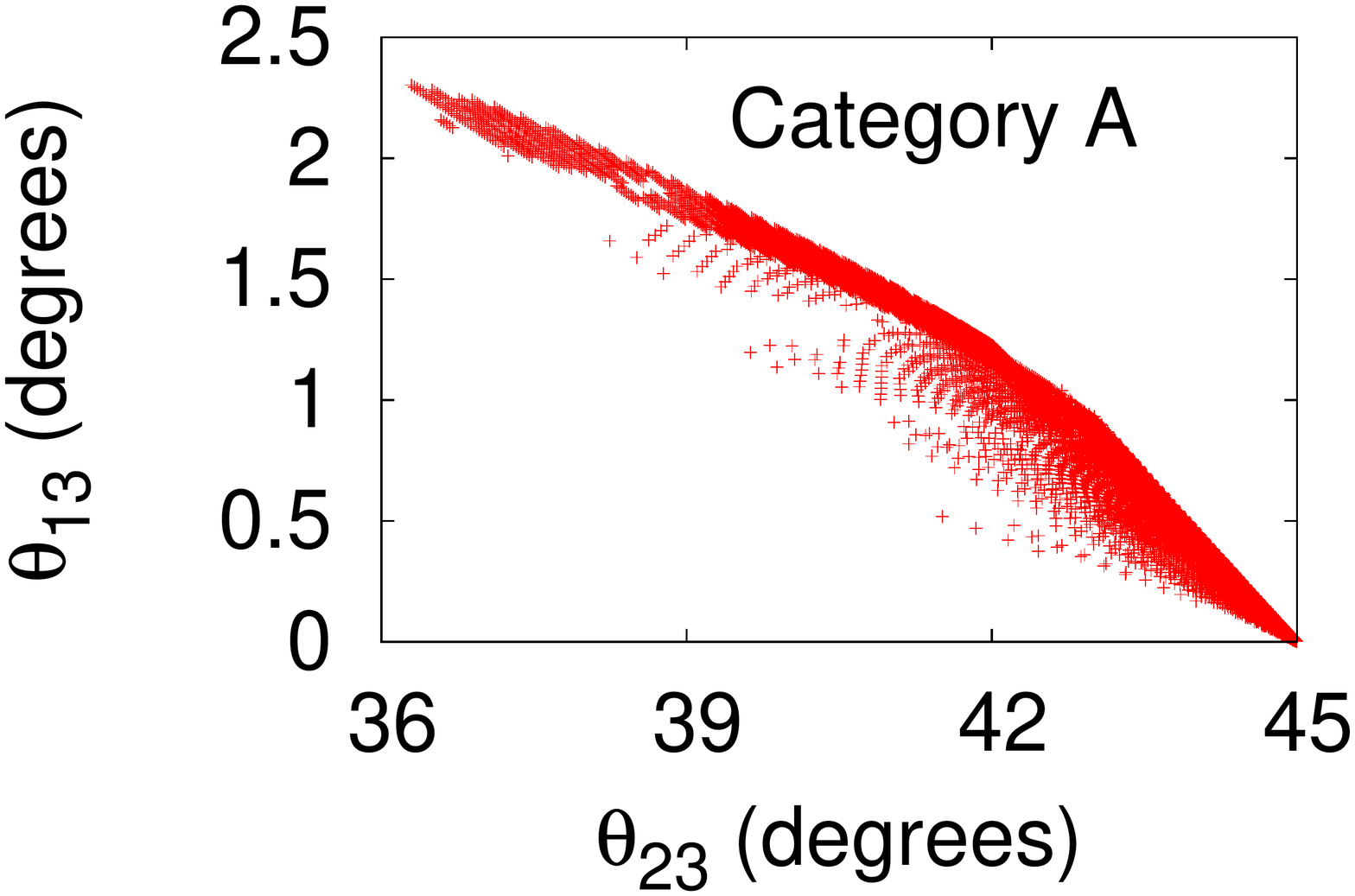}
\includegraphics[width=3.5cm,height=3.5cm]{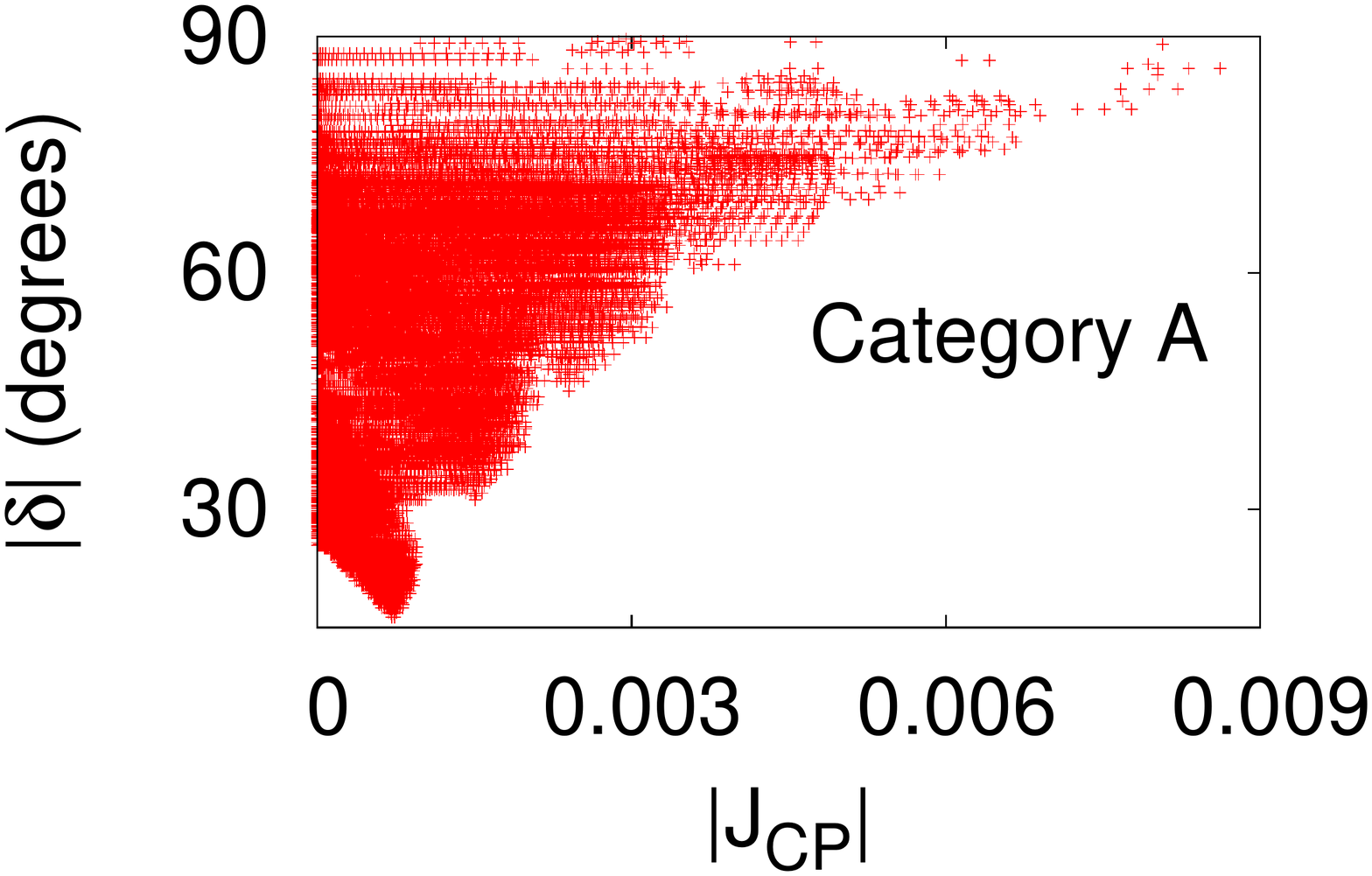}
\vskip 0.1cm
\includegraphics[width=3.5cm,height=3.5cm]{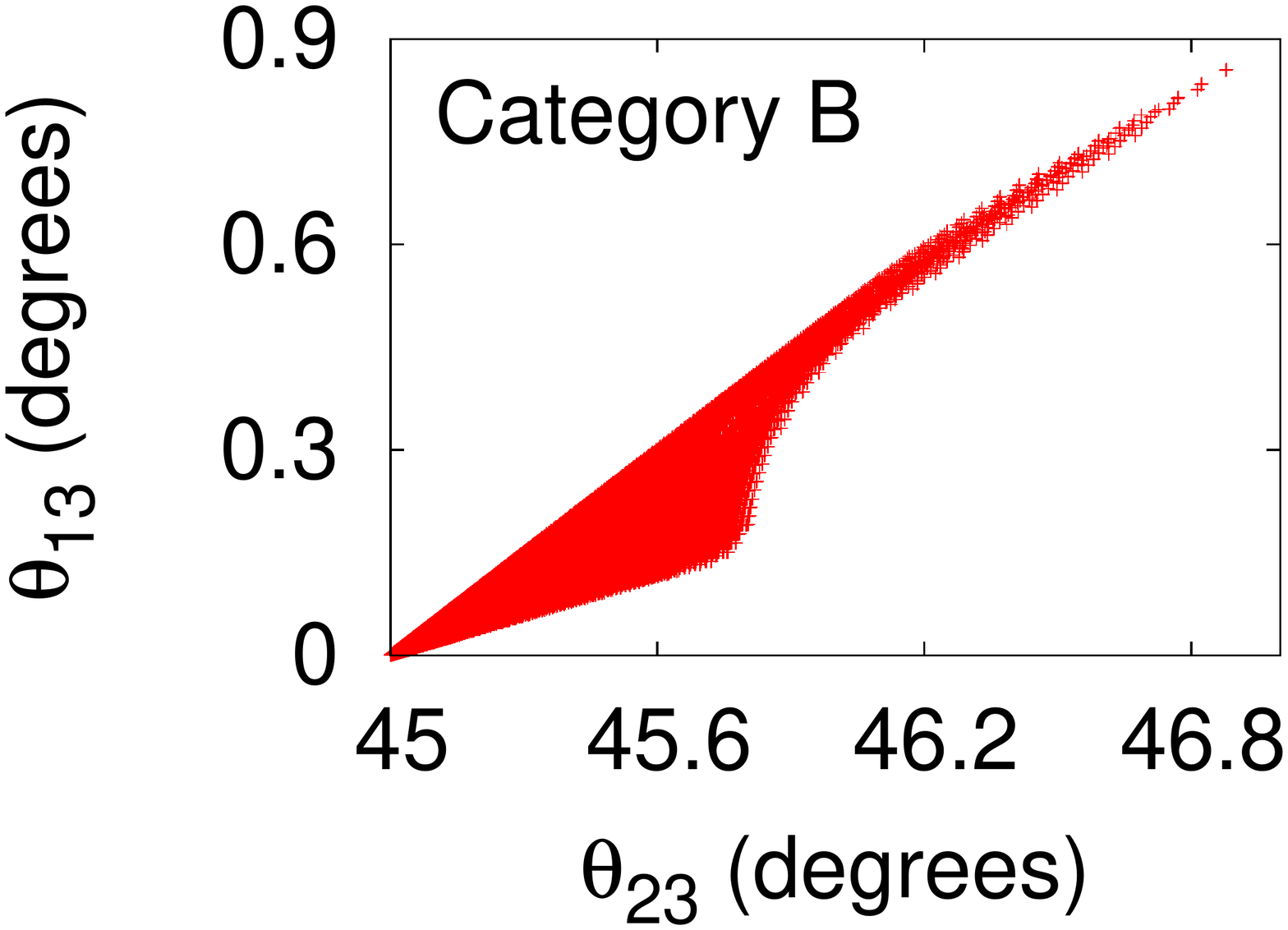}
\includegraphics[width=3.5cm,height=3.5cm]{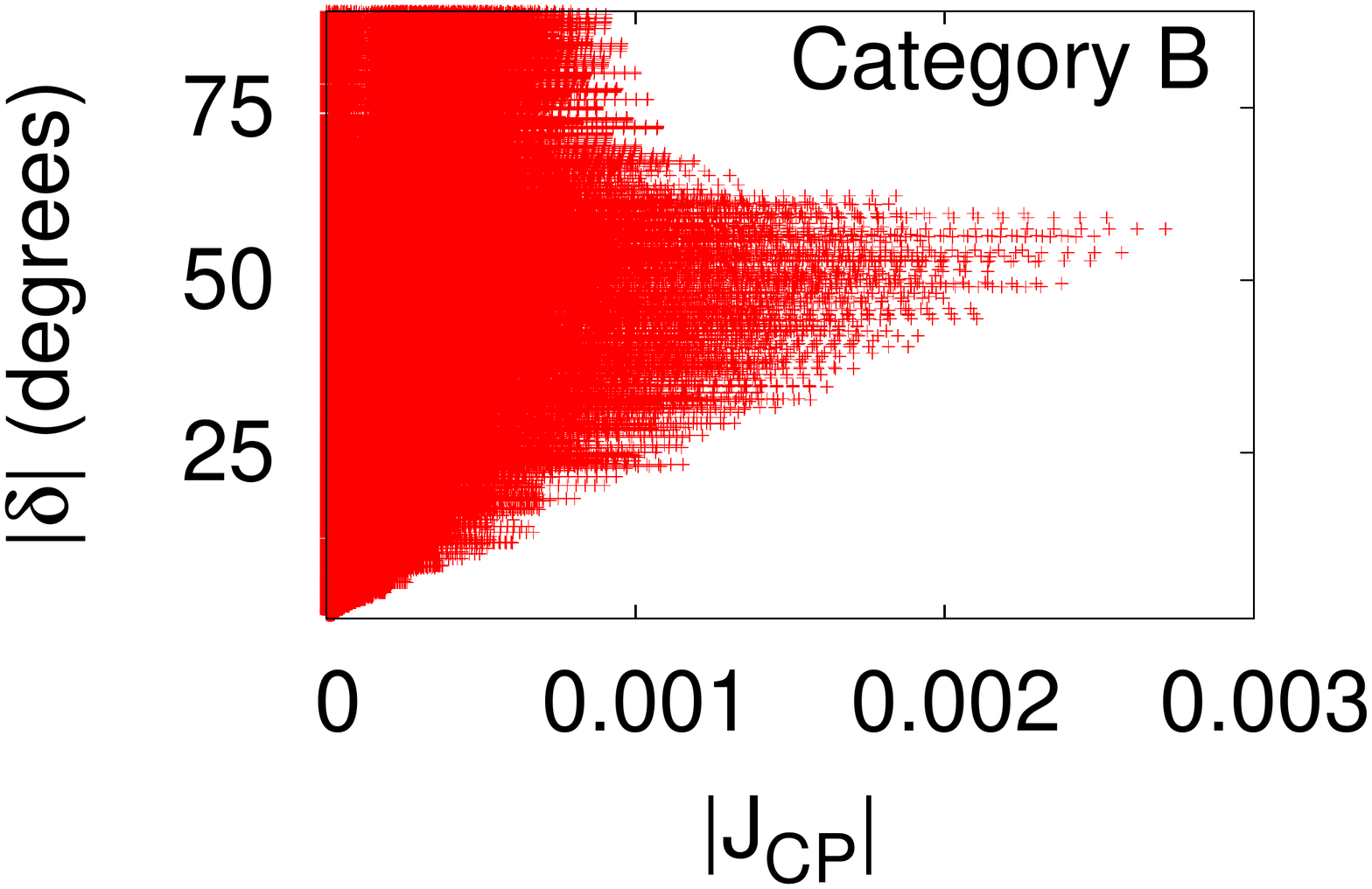}
\caption{\small{{(Color online) Allowed $\theta_{13}$ vs 
$\theta_{23}$ (left) and $|\delta|$ vs 
$|J_{CP}|$ (right) for Category $A$ (top) and Category $B$ (bottom).}}}
\end{flushleft}
\end{figure}
to place zeroes in any number of fermionic mass matrix elements was proposed 
\cite{31,311}
some time ago. A lot of work \cite{33} has in the interim 
been done with the flavor symmetry groups 
$A_4$ and $S_3$, though we are unaware of any model that predicts our 
$m_D$'s.
While there are possibly fruitful directions of interest,  
a credible understanding of the origin of the 
present four zero Yukawa textures remains a theoretical challenge.
\par
To sum up, from allowed $\mu\tau$ symmetric four zero Yukawa textures, 
we have obtained explicit values of the three light neutrino masses in terms 
of $\sqrt{\Delta_{21}^2}$. The predicted magnitude of their sum 
will soon be tested cosmologically. According to our scheme, 
$0\nu\beta\beta$ decay may or may not be 
observed in the near future, but CP violation, to be
measured in long baseline experiments at neutrino factories, 
will tell us how it 
effected leptogenesis in the early Universe.  
\par
{\sl Acknowledgements}:
We thank A.~Dighe for discussions on the cosmological bound. P.~R. was 
supported by a DAE Raja Ramanna fellowship.


\begin{thebibliography}{99}
\bibitem{new} The phases of $a_2/b_1$ in Category $A$ and of $a_1/b_2$ in 
Category $B$ have been 
absorbed in the $\nu_e$ field.
\end{thebibliography}
\end{document}